\documentclass[lettersize,journal]{IEEEtran}
\usepackage{amsmath,amsfonts}
\usepackage{algorithmic}
\usepackage{algorithm}
\usepackage{array}
\usepackage[caption=false,font=normalsize,labelfont=sf,textfont=sf]{subfig}
\usepackage{textcomp}
\usepackage{stfloats}
\usepackage{url}
\usepackage{hyperref}
\usepackage{orcidlink}
\usepackage{verbatim}
\usepackage{graphicx}
\usepackage{cite}
\usepackage[margin=0.54in]{geometry}
\begin{document}
    \title{Image Synthesis Using Spintronic Deep Convolutional Generative Adversarial Network}
    \author{Saumya Gupta\orcidlink{0009-0001-7636-1553}, Abhinandan\orcidlink{0009-0006-0570-5212}, Venkatesh vadde\orcidlink{0009-0001-4874-7563}, Bhaskaran Muralidharan\orcidlink{0000-0003-3541-5102}, Abhishek Sharma\orcidlink{0000-0002-3635-3081}
    \thanks{Saumya Gupta, Venkatesh vadde and Bhaskaran Muralidharan (email:bm@ee.iitb.ac.in) are with the Department of Electrical Engineering, IIT Bombay, Mumbai 400076, India}
    \thanks{ Abhinandan is with Department of Electrical Engineering, IIT
    Ropar, Rupnagar, Punjab 140001, India}
    \thanks{ Abhishek Sharma is with the School of Computing and Electrical Engineering (SCEE), IIT
    Mandi, Himachal Pradesh, 175005, India (e-mail: abhi@iitmandi.ac.in).}
    \thanks{The author BM acknowledges funding from the the Inani Chair Professorship fund through Grant No. DO/2024-INAN/001-001. The author SG acknowledges the financial support from the University Grants Commission (UGC). The author, AS, acknowledges the support provided by SERB, Grant No. SRG/2023/001327.}}
    \maketitle
    \begin{abstract}
   The computational requirements of generative adversarial networks (GANs) exceed the limit of conventional Von Neumann architectures, necessitating energy efficient alternatives such as neuromorphic spintronics. This work presents a hybrid CMOS-spintronic deep convolutional generative adversarial network (DCGAN) architecture for synthetic image generation. The proposed generative vision model approach follows the standard framework, leveraging generator and discriminators adversarial training with our designed spintronics hardware for deconvolution, convolution, and activation layers of the DCGAN architecture. To enable hardware aware spintronic implementation, the generator’s deconvolution layers are restructured as zero padded convolution, allowing seamless integration with a 6-bit skyrmion based synapse in a crossbar, without compromising training performance. Nonlinear activation functions are implemented using a hybrid CMOS domain wall based Rectified linear unit (ReLU) and Leaky ReLU units. Our proposed tunable Leaky ReLU employs domain wall position coded, continuous resistance states and a piecewise uniaxial parabolic anisotropy profile with a parallel MTJ readout, exhibiting energy consumption of 0.192 pJ. Our spintronic DCGAN model demonstrates adaptability across both grayscale and colored datasets, achieving Fréchet Inception Distances (FID) of 27.5 for the Fashion MNIST and 45.4 for Anime Face datasets, with testing energy (training energy) of 4.9 nJ (14.97~nJ/image) and 24.72 nJ (74.7 nJ/image). 
    \end{abstract}
\section{Introduction}\label{sec:introduction}
\IEEEPARstart{G}{enerative} models represent a cutting edge class of of deep learning techniques within the realm of artificial intelligence\cite{balasubramaniam2024road}, enabling the synthesis of data that closely resembles real world distributions. Among these, diffusion models\cite{dhariwal2021diffusion} have recently emerged as a formidable competitor in image generation tasks. In parallel, generative adversarial Networks (GANs)\cite{goodfellow2020generative} remain a widely adopted and computationally efficient generative framework, particularly well suited for fast\cite{dhariwal2021diffusion} and real time applications such as prototyping, video games, and style transfer\cite{long2024sketchar}. 
GANs function primarily
as unsupervised learning framework, wherein a generator network  possess the unique ability to generate
novel instances from original datasets, while the discriminator network autonomously discerns and internalize patterns or regularities inherent in the input data. For image centric task, Deep Convolutional GANs (DCGANs)\cite{radford2015unsupervised} extend the GAN framework by incorporating convolutional architectures to effectively capturing spatial dependencies. DCGAN replaces pooling with strided convolution in the discriminator, and employs deconvolution in the generator. Further, the use of rectified linear unit (ReLU) and leaky ReLU activations mitigates the dying ReLU problem, enabling improved gradient flow, enhanced training stability, and higher image synthesis quality. The architectural guidelines provide a balanced mini max model less prone to hyperparameter than GAN.\\
\indent Hardware acceleration is a critical design consideration for GAN due to the computational and energy demand for real time and edge applications. Conventional GAN hardware implementations predominantly rely on CMOS technology
\cite{TatsuyaKaneko2019}, leveraging optimized application specific accelerators and network level optimizations across FPGA\cite{yin2021reconfigurable}, ASIC\cite{xu2020reconfigurable} and  edge TPU\cite{xu2020accelerating} platforms. However, these systems still face a mismatch between data movement and computation\cite{peccerillo2022survey}, motivating the exploration of alternative technologies\cite{guo2025high} to handle the deep neural network operations efficiently\cite{chen2019efficient}. Processing-in-memory (PIM) architectures based on resistive RAM (RRAM or ReRAM)\cite{chen2019cmos} offer fast and in memory computations but suffer from limited bit precision and substantial peripheral circuit overheads\cite{rakin2018pim}.\\
\indent Spintronic neuromorphic hardware has emerged as a promising alternative due to its CMOS compatibility\cite{chung20164gbit}, high endurance, intrinsic oscillatory, plastic, linear, and stochastic behaviours\cite{vadde2024spintronic}, along with significantly reduced time and energy requirements. A limited number of studies have explored spintronic hardware implementations inspired by generative adversarial learning paradigms. Existing efforts primarily leverage magnetic tunnel junction (MTJ) based primitives, including SOT-MRAM for processing-in-memory acceleration \cite{rakin2018pim}, superparamagnetic MTJs \cite{koh2025closed} or MTJ-based random number generators \cite{bao2025securing}. While these works demonstrate the feasibility of employing spintronic devices for components related to generative adversarial learning, they do not provide a comprehensive, end-to-end DCGAN realization encompassing device to system level evaluation. Beyond these GAN inspired implementations, spintronic device primitives tailored for neural computation are reported in literature such as skyrmion-based synapses for weighted summation and domain wall (DW) devices for neuron or activation like functionality highlighting the potential of magnetic quasi particles for CNN, SNN, and ANN architectures\cite{das2023bilayer}.\\
\indent Building on these advancements, we leverage the inherent advantages of spintronics to design specialized hardware modules using skyrmion and domain wall for DCGAN layers, enabling efficient and scalable image generation. This work extends our earlier design of a skyrmion based synapse and domain wall based rectified linear unit (ReLU)\cite{gupta2025magnetic}. Here, we propose a hybrid CMOS DW based tunable leaky ReLU device tailored for DCGAN requirements. In our architecture, skyrmion based synapses realize the convolution operations in the discriminator and the deconvolution operations in the generator, while hybrid CMOS DW devices implement the ReLU and leaky ReLU activations essential for adversarial image synthesis.\\
\indent The paper is organized as follows: In Sec.\ref{dcgan}, we presents our DCGAN implementation with a modified generator and integration of skyrmion based synapse with domain wall based ReLU and Leaky ReLU units. Sec.\ref{Spintronic devices for DCGAN} details the design model for all simulated and proposed spintronic hardware. Sec.~\ref{methodology} details the dataset, device simulation setup, network architecture and training methodology. Sec.~\ref{results} reports device level and network level results, including loss evolution, Fréchet Inception Distance (FID), noise evaluation, energy consumption, quantization behavior, and hardware constraints. Finally, Sec. \ref{conclusion} concludes the paper.
\section{DCGAN Implementation}\label{dcgan}
\subsection{Principle of DCGAN}
\begin{figure*}[!t]
\centering
\includegraphics[width=1\textwidth,height=3in]{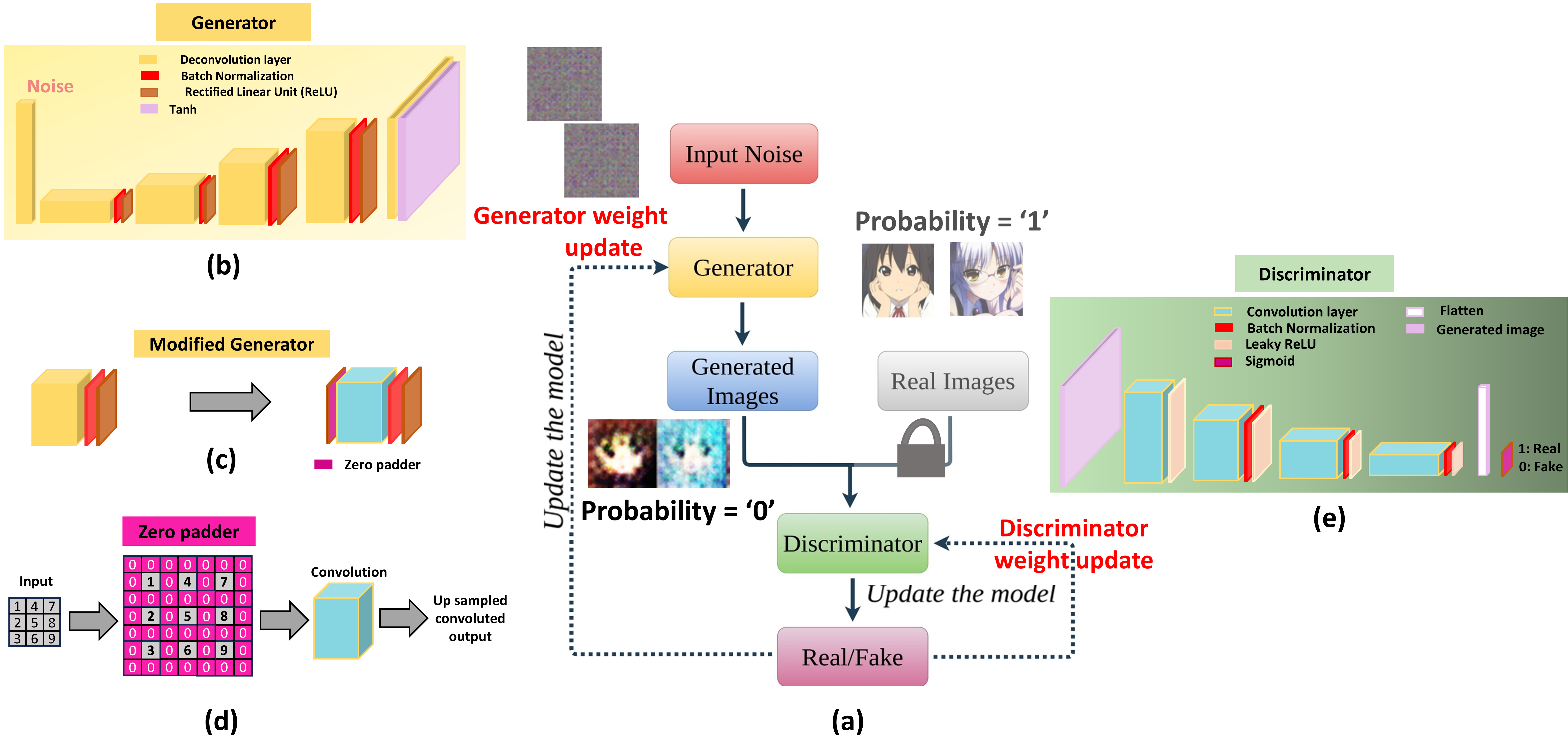}
\caption{(a) Block Diagram of Generative Adversarial Network (b) DCGAN generator block diagram (c) Modified DCGAN generator (d) Deconvolution done with zero padding plus convolution (e) DCGAN discriminator block diagram.}
\label{Block Diagram}
\end{figure*}

\indent The GAN framework (Fig.~\ref{Block Diagram}(a)) comprises of two distinct neural networks: generator and the discriminator that engage in dynamic interaction characterized by an adversarial process.
\noindent
 The discriminator network discerns between real and synthetic data. While, the generator network produces artificial data across various modalities such as text, audio, or images. During training, generator strives to fool discriminator, while the discriminator refines its ability to discriminate synthetic data, jointly enhancing GAN performance \cite{goodfellow2020generative}. Several GANs architectures (Vanilla GAN, BigGAN, StyleGAN etc.\cite{durgadevi2021generative}) share this common generator and discriminator design feature. In this work, we employ DCGAN architecture, leveraging deep CNN as the foundational component due to their strong spatial modeling capability in image generation.\\
\indent The DCGAN generator network (Fig.~\ref{Block Diagram}(b)) transforms input noise vector using deconvolution layers. Then Batch normalization stabilizes training with normalized input having zero mean and unit variance, catalyzing onset of model learning to avoid mode collapse. Further, the batch normalized output is fed to ReLU that introduces nonlinearity before the final tanh output layer, to ensure pixel values between [-1 1].\\
\indent We modified the generator's  deconvolution layer as a zero padded plus convolution layer to upsample noise into a complex image.  Our modified DCGAN generator layer as shown in Fig.~\ref{Block Diagram}(c) performs convolution with our skyrmion based synapse in a crossbar array. The structural changes in generator does not jeopardize the integrity of the training framework as a whole (see Sec.~\ref{modi_gen}). We also simulated ReLU with our hybrid CMOS DW activation circuit as discussed in Sec. \ref{rlr}.\\
\indent The DCGAN discriminator (see Fig.~\ref{Block Diagram}(e)) is a binary classifier network built using convolution, followed by batch normalization and Leaky ReLU, which mitigates the dying ReLU problem with added non-linearity. We propose a hybrid CMOS DW based leaky ReLU design for the discriminator that can be tuned as per the network requirements (see  Sec. \ref{rlr}). Strided convolution layer in the discriminator extracts hierarchical features by skipping pixels as the kernels slide across the input and perform matrix vector multiplication. On traditional Von Neumann hardware, the matrix vector multiplication requires extensive memory access, whereas crossbar arrays offer efficient in-memory alternative. Crossbar arrays store kernel weights as synaptic conductances along vertical lines, while inputs are applied horizontally. The resulting column currents naturally perform the weighted sum, reducing latency and energy. Skyrmion-based synapses\cite{gupta2025magnetic} provide nanoscale footprint, non-volatility, topological stability, low-current tunability, and fast current-driven dynamics, making them ideal for continuous tunable convolution weights in crossbar architectures\cite{jung2022crossbar}. The discriminator's final sigmoid layer outputs probabilities [0 : Fake, 1 : Real].\\
\indent Generator weights are unaffected during the discriminator training as shown in Fig.~\ref{Block Diagram}(a). While, discriminator plays an essential role in generator training as the feedback optimizes generator output. This adversarial iteratively improves both networks, producing realistic synthetic images (Sec.~\ref{Evaluation metrics}).
The specialized spintronic circuits: skyrmion based synapse for convolution and deconvolution layers, and hybrid CMOS domain wall activation functions for ReLU and leaky ReLU are presented for the hardware implementation of DCGAN, described in the following Sec.~\ref{Spintronic devices for DCGAN}.
\subsection{Modified Generator}\label{modi_gen}
\indent In DCGAN, the generator network removes fully connected layers and replace them with deconvolution. Deconvolution layer modified as zero padding plus convolution, upsamples the feature map by inserting empty pixels between samples, creating a one-to-many mapping that increases spatial resolution. As shown in Fig.~\ref{Block Diagram}d, individual input elements are augmented by adding zeros along both rows and columns, and then convolved to produce an upsampled matrix. While this approach introduces redundant zero operations, it simplifies hardware design and enables direct integration of the skyrmion-based synapse into crossbar arrays for efficient hardware level image generation.
\section{Spintronic devices for DCGAN}\label{Spintronic devices for DCGAN}
\indent The spintronic DCGAN implementation for image synthesis task includes our previous work on 6-bit skyrmionic synapse and the hybrid CMOS DW ReLU\cite{gupta2025magnetic} as mentioned in the introduction. In this work, we extend the framework by presenting a hybrid CMOS DW based leaky ReLU design and integrating into a complete DCGAN pipeline. The architecture supports convolution, deconvolution, and activation functions (ReLU and leaky ReLU) using skyrmion and domain wall devices detailed in the following subsections.
\subsection{Skyrmion based synapse}\label{syn}
\begin{figure*}
    \centering
    \includegraphics[height= 2.8in,width=0.95\linewidth, height=3.2in]{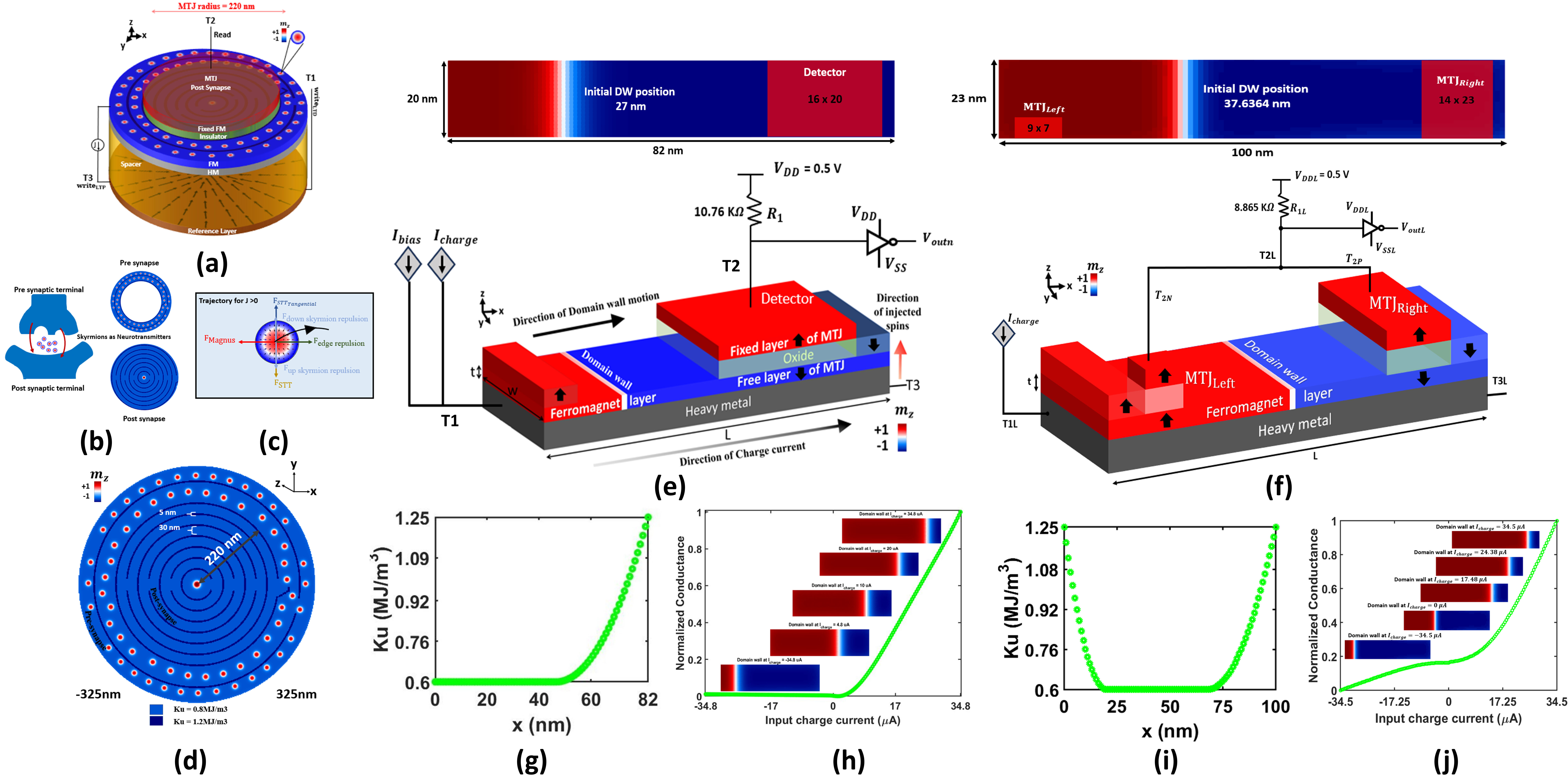}
    \caption{(a) 3D schematic of the 6-bit skyrmion synapse device. (b) Biological neural analogy. (c) Skyrmion forces under applied current density. (d) 2D view : FM layer with 64 skyrmions and labyrinthine uniaxial anisotropy profile with outer and inner rings that form the pre synapse region (magnetization direction: blue = into the plane, white = in-plane and red = out of the plane). (e) 2D domain wall track with detector and 3D CMOS domain wall ReLU circuit. (f) 2D domain wall track with left detector (MTJ\textsubscript{Left}) and right detector (MTJ\textsubscript{Right}),and 3D CMOS domain wall Leaky ReLU circuit. (g) Parabolic uniaxial anisotropy~(K\textsubscript{u}) profile for ReLU. (h) Normalized conductance versus input charge current with domain wall snapshots for ReLU. (i)~Piecewise parabolic uniaxial anisotropy~(K\textsubscript{u}) profile (PMA) for leaky ReLU device. (j) Normalized conductance versus input charge current with domain wall snapshots for Leaky ReLU.}\label{Fig.3}
    \label{lrelu and leakyrelu design}
\end{figure*}

\indent The 6-bit circular skyrmionic synapse\cite{gupta2025magnetic} (Fig.~\ref{lrelu and leakyrelu design}(a)) consist of a reference FM layer with fixed vortex-like magnetization (RL) and a free FM layer separated by a heavy metal layer inducing interfacial Dzyaloshinskii-Moriya Interaction (DMI). The device mimics biological neural network with a 220 nm post-synapse and a pre-synapse between 220 to 325 nm region, where skyrmions act as neurotransmitters (Fig.~\ref{lrelu and leakyrelu design}(b)). A high Ku (1.2 MJ/$\mathrm{m^3}$) anisotropy ring forms a 30 nm spaced labyrinthine track that hosts 64 skyrmions (32 inner, 32 outer) within the 325 nm radius (Fig.~\ref{lrelu and leakyrelu design}(d)), achievable using $\mathrm{He^{+}}$ ion irradiation\cite{juge2021helium}. Skyrmions are nucleated and injected into the pre-synapse using a $\mathrm{10^4~MA/cm^2}$ for 2 ns with +z spin polarized current pulse, followed by 10 ns relaxation and read via magnetoresistance enabled by the MTJ.\\
\indent The in-plane write current (T3 to T1) induces spin torque forming a vortex-like spin polarization\cite{jiang2018dynamics}. Depending on the injected current direction (+z or –z), the skyrmion lattice gyrates clockwise or anticlockwise along high Ku constrictions as it enters or exits the MTJ. Positive current (negative current) pulses drive skyrmions from the outer ring to inner ring and then post-synapse (post synapse to outward rings or pre synapse), increasing (decreasing) conductance. Conductance varies from $\mathrm{G_{min}}$ to $\mathrm{G_{max}}$ according to the number of skyrmions under the detector. The device supports all synaptic behaviors including initiation: time to reach detector; Long term Potentiation (LTP): linear conductance rise under positive pulses; Long term Depression (LTD): linear decrease under negative pulses; Short Term Plasticity
(STP): intermediate stable state, and reset: re-spacing skyrmions in the pre-synapse. The linear conductance encodes synaptic weights.
\subsection{CMOS hybrid domain wall ReLU and Leaky ReLU}\label{rlr}
Figure~\ref{lrelu and leakyrelu design}(e) and \ref{lrelu and leakyrelu design}(f) illustrate the simulated hybrid CMOS DW ReLU\cite{gupta2025magnetic} and our proposed leaky ReLU device, respectively. The proposed DW based leaky ReLU device shares the same underlying physical structure as the ReLU device, comprising a SOT driven monolayer FM with a single domain wall and a perpendicular magnetic anisotropy (PMA) profile achieved via controlled oxidation process under bias\cite{fassatoui2021kinetics}. Device dimensions, detector placement, MTJ readout configuration and uniaxial anisotropy profile (Ku) is selected differently to realize both ReLU and leaky ReLU characteristics within the same device setup. Both devices incorporate a CMOS inverter circuit and a single domain wall in FM layer. The leaky ReLU has FM dimension 100 nm x 23 nm x 1 nm, featuring PMA with two oppositely magnetized regions separated by a DW.\\
\indent In our Leaky ReLU design, the FM layer uses a piecewise parabolic Ku profile with two distinct slopes to obtain non-zero value (ax) for negative axis. Figure~\ref{lrelu and leakyrelu design}(i) shows the parabolic Ku variation in the range 0.6 $\mathrm{MJ/m^3}$ to 1.25 $\mathrm{MJ/m^3}$ in 0 nm to 21 nm and 68 nm to 100 nm regions with different slopes and remains constant (0.6$\mathrm{MJ/m^3}$) between 21 nm to 68 nm. The steeper Ku gradient (left side) results in smaller conductance changes for negative currents compared to shallower right side gradient with faster and larger conductance changes for positive currents (see Fig.~\ref{lrelu and leakyrelu design}j). The DW is initially relaxed at 37.6364 nm and driven by electrical charge current from -34.5 $\mathrm{\mu} A$ to 34.5 $\mathrm{\mu}A$. The two parabolic regions induce reduced DW velocity for both applied current polarities. A constant Ku with a 20 nm to 30 nm gap between regions ensure smooth and stable DW motion.
Both devices self-reset within a few nanoseconds without external fields\cite{siddiqui2019magnetic} or auxillary circuits\cite{liu2021controllable,brigner2019shape}. The physical principles of DW motion remain the same for both our devices. At the FM-HM interface the spin orbit coupling (SOC) leads to DMI interaction, stabilizing the neel DW. The presence of strong DMI rotates the DW moment as well as tilts the DW line profile as a result of energy minimization on the x-y axis.\\
\indent The in-plane charge current produces a spin polarized current in z direction (-y polarized), driving the DW beneath the detector region defined by the spatially varying ku profile.
The MTJ readout (free FM layer with DW, oxide barrier and fixed FM layer) converts the domain wall position modulated by $\mathrm{\mathrm{I_{charge}}}$ (at terminal T1 or T1L) to conductance values.\\
\indent For the ReLU device, a single MTJ detector placed between 63 nm to 79 nm ('read' terminal T2 to T3) senses DW motion only for positive current in the presence of a bias current, producing a linear conductance increase. For negative current, the DW remains outside the readout region (towards -x), resulting in a nearly zero conductance. Figure~\ref{lrelu and leakyrelu design}(h) shows the domain wall position corresponding to the applied $\mathrm{I_{charge}}$ on the FM nano track.\\
\indent For the leaky ReLU device, a parallel MTJ readout (from read terminal T2L to T3L), sums conductances from the left ($T_{2N}$) and right ($T_{2P}$) MTJs. The left MTJ (9 nm x 7 nm) is  placed from 4 nm to 13 nm in length and 1 nm to 7 nm in height. The right MTJ is comparatively larger (14 nm x 23 nm),  placed between 83 nm to 97 nm in length and 1 nm to 23 nm in height. Summed conductance increases monotonically for both current polarities with different slopes, achieving the leaky ReLU behavior without additional bias current. Optimized MTJ dimensions and placement ensures the conductance slopes converge at zero current, yielding a continuous transfer characteristic. Notably, only one CMOS inverter is required, identical to the ReLU, enabling energy efficient operation. A 3 nm exchange bias layer on both the FM ends prevent DW annihilation at the boundaries\cite{kaushik2020comparing}. The derived resistances from respective devices feed a resistor $R_{1}$ for ReLU (and $R_{1L}$ for leaky ReLU), and a CMOS inverter pair to produce desired non-linear activation functions ReLU (Fig.~ \ref{fig_sim}(d)) and leaky ReLU (Fig.~ \ref{fig_sim}(e)) respectively. 
\section{Proposed methodology}\label{methodology}
Fig.~\ref{fig_sim} presents the complete simulation methodology as a schematic, including the device block and results. Figure~\ref{fig_sim} (a) and (c) correspond to device simulation blocks for synapse and ReLU/Leaky ReLU device respectively. The simulated device characteristics are integrated into the DCGAN training framework (see Fig.~\ref{fig_sim}(f)), implemented in Python using the PyTorch library. 
\subsection{Datasets}
Two datasets were utilized for the network training: i) Fashion-MNIST dataset and  ii) Anime dataset. We performed the spintronic DCGAN training on both the datasets to create adversarial images,  demonstrating our model’s adaptability across both grayscale and colored datasets.
\subsubsection{Fashion-MNIST dataset}
Fashion-MNIST is a grayscale image dataset of 28 × 28 pixels containing ten categories of Fashion items. The dataset (by Zalando Research) contains a total of 70,000 images, 60,000 training images and 10,000 testing images. 
\subsubsection{Anime dataset}
The Anime Face Dataset (by splcher)\cite{churchill2019anime} is a real-world dataset containing 63,565 RGB Anime face images with resolution between 
25 × 25 to 220 × 220 pixels ( average $\sim$ 89.6 × 89.6 pixels). For our experiments, we used 57,209 training images and 6,356 testing images. All images were resized to 64 × 64 pixels to ensure consistency and enable our DCGAN to generate visually realistic and diverse anime style facial images.

\subsection{Network Architecture}
\indent Generator Architecture: The generator takes a 1-D latent vector of size 100 and progressively upsamples it using first transposed-convolution layer followed by four zero padded plus convolution layer.
The channel progression is dataset dependent. Then Batch Normalization is implemented followed by ReLU activation after each layer except the output, which uses Tanh. \\
\indent Discriminator Architecture: The discriminator processes an input image using four convolutional layers with feature maps {64 $\rightarrow$ 128 $\rightarrow$ 256 $\rightarrow$ 512}.
Each layer is followed by Batch Normalization and Leaky ReLU (a = 0.2) except the output, which uses sigmoid activation to output the real/fake probability.\\
\indent For Fashion MNIST dataset, Generator outputs 28×28 images with channel progression {256 $\rightarrow$ 128 $\rightarrow$ 64 $\rightarrow$ 32 $\rightarrow$ 16 $\rightarrow$ 1}. The discriminator processes 28 × 28 inputs.
For Anime face dataset, Generator outputs 64 × 64 images with channel progression {512 $\rightarrow$ 256 $\rightarrow$ 128 $\rightarrow$ 64 $\rightarrow$ 3}. The discriminator processes 64 × 64 inputs.

\subsection{Synapse}

The synapse is simulated in the micromagnetic simulation platform,  OOMMF(added DMI extension module)\cite{donahue1999oommf} using Co-Pt\cite{sampaio2013nucleation} parameters at room temperature. The 6 bit circular device (650 nm x 650 nm x 0.5 nm), with a constant uniaxial anisotropy = 0.8$\mathrm{MJ/m^3}$ and a high Ku of 1.2$\mathrm{MJ/m^3}$. Refer to \cite{gupta2025magnetic} for full micromagnetic simulation parameters.
Skyrmion dynamics follow Landau Lifshitz Gilbert Slonczewski (LLGS) and thiele, where  steady motion results from balancing Magnus and boundary forces. Conductance is computed using the extended Julliere \cite{julliere1975tunneling} and Slonczewski \cite{slonczewski1989conductance} models, normalized using NEGF.\\
\indent The energy dissipated per weight update is given by:
 \begin{equation}
     E_{write} = I^{2}_{c}R_{write}T_p
 \end{equation}
\indent where, $\mathrm{I_c}$ is the charge current, $\mathrm{T_p}$ is the pulse duration, and $\mathrm{R_{Write}}$ is the valet-fert bilayer resistance\cite{gupta2025magnetic}. The synapse simulation blocks is shown in Fig.~\ref{fig_sim}a. 

\subsection{CMOS hybrid domain wall}
We perform the micro-magnetic simulations for the DW based ReLU and leaky ReLU on a GPU accelerated numerical package (mumax3). The cell size is 1 nm x 1 nm x 1 nm. The mumax3 uses custom field functionality to implement SOT with LLG equation. The resistance of the HM is given by $\mathrm{R=\rho l_{HM}/W_{HM}t_{HM}}$, where $\mathrm{l_{HM}}$ is the length of HM, $\rho$ is the resistivity of HM~($\mathrm{Au_{0.25}Pt_{0.75}}$), $\mathrm{t_{HM}}$ is the thickness of HM and $\mathrm{W_{HM}}$ is the width of HM. The spin current from the heavy metal layer is given by Hirsch, Takahashi, and Maekawa.
where $\mathrm{I_s, \theta, l_{FM}, t_{HM}, I_c}$, and P are the magnitude of spin current, spin Hall angle, length of the FM, thickness of the HM layer, charge current and polarization of the spin current respectively\cite{gupta2025magnetic}.\\
\indent The device is simulated using $\mathrm{Co-Au_{25}Pt_{75}}$\cite{yang2023magnetic} system with following parameters: Saturation magnetization ($M_{s}$) = 580 KA/m, Gilbert damping factor ($\alpha$) = 0.3, DMI constant ($D$) = 3 $\mathrm{mJ/m^2}$, Exchange stiffness constant ($A_{intra}$) = 15 pJ/m, Spin polarization factor ($P$) = 0.614, $\frac{|field-like torque|}{|damping-like torque|}$ ($\xi$) = 0.2, Spin Hall Angle ($\theta$) = 0.3, resistivity of HM ($\rho_{HM}$) = 83 $\mathrm{\mu\Omega cm}$, Resistance of HM ($R_{HM}$) = 850.75 $\Omega$, Thickness of HM ($t_{HM}$) = 4 nm, MTJ Capacitance ($C_{MTJ}$) = 26.562 aF, constant uniaxial anisotropy ($ Ku_c$) = 0.6 $\mathrm{MJ/m^3}$. The conductance calculation for DW device is same as that of synapse. Further, the obtained resistance values are embedded in verilog A and the circuit is simulated in HSPICE. The ReLU and Leaky ReLU simulation blocks is shown in Fig.~\ref{fig_sim}(c).The corresponding additional parameters for ReLU and leaky ReLU are mentioned below:
\subsubsection{ReLU} \label{relu}
The DW device uses a heavy metal layer with  82 nm x 20 nm dimension, and FM Volume = 1640 $\mathrm{nm^3}$. Refer to \cite{gupta2025magnetic} for full ReLU circuit parameters.
\subsubsection{Leaky ReLU}\label{leakyr}
The DW device has length ($\mathrm{L_{HM}}$) and Width ($\mathrm{W_{HM}}$) of HM as 100 nm and 23 nm respectively. The Volume of FM is = 2300 $\mathrm{nm^3}$.  The CMOS hybrid circuit is implemented with following parameters: Reference Resistor ($R_{1L}$) = 8.865 K$\Omega$, No biasing current is required, Simulation time step ($\Delta_t$) = 0.5 ps, Voltage sources ($V_{DD}$, $V_{SS}$) are 0.5 V, -0.165 V respectively. Also, for the CMOS inverter pair, transistor sizing ratios for nmos $\mathrm{{W_n}/{L_n}}$ and pmos $\mathrm{{W_p}/{L_p}}$ are 2.2/1 and 5.1/1 respectively.
The left parabolic Ku profile between 0 nm to 21 nm is governed by $\mathrm{Ku= Ku_c + 1625*(21-i)^2}$, and for the right parabolic ku profile, $\mathrm{Ku= Ku_c + 634.766*(i-68)^2}$, where i represents the distinct regions varying in their respective spaces.

\subsection{Training of DCGAN}\label{Training of DCGAN}

The DCGAN training uses Binary Cross Entropy (BCE) loss for both the discriminator and generator network. To compute the generator's loss, we generate a batch of synthetic images, pass them through the discriminator, and assign the target label as 1 (real). Using the feedback, the BCE loss drives the generator to produce images that the discriminator classifies as real.\\
\begin{figure*}[!t]
\centering
\includegraphics[width=0.8\linewidth,height=3in]{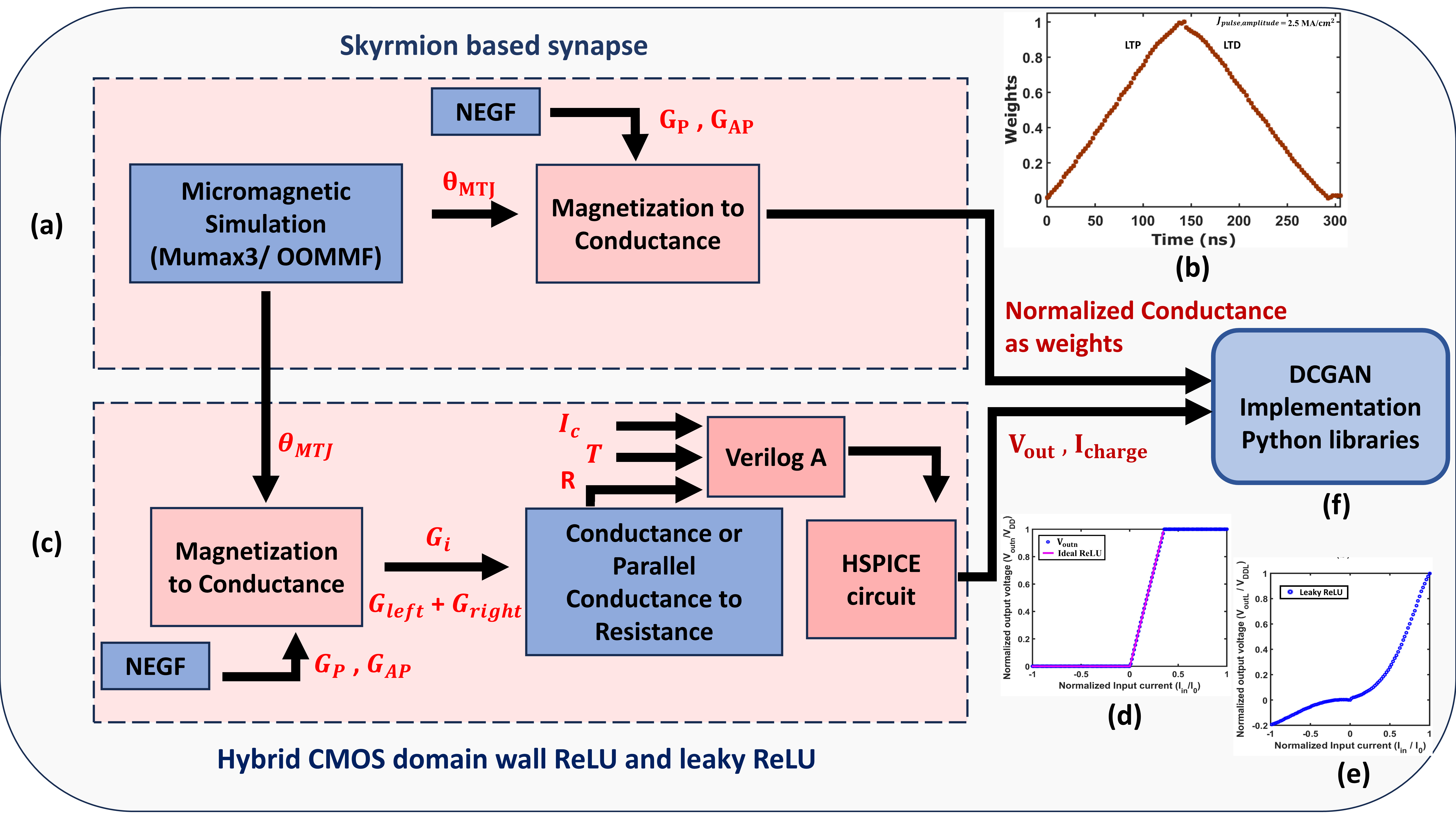}
\caption{(a) Simulation flow for Skyrmion based synapse. (b) Synaptic weight versus simulation time. (c) Simulation flow for ReLU and Leaky ReLU. (d)~ReLU function. (e) Leaky ReLU function realized using a hybrid CMOS domain wall device. (a,e,f) Overall simulation methodology.}
\label{fig_sim}
\end{figure*}
\indent The training parameters for Fashion MNIST dataset (Anime face dataset) are as follows: epochs = 300 (80), batch size = 128, latent size = 100, optimizer = Adam, image size = 28 (64), learning rate of generator = 2e-4 (1e-4) and learning rate of discriminator = 2e-4.
\section{Discussion}\label{results}
\subsection{Device results}
\subsubsection{synapse}
In the synapse device (Sec.~\ref{syn}) conductance varies proportionally with the number of skyrmions, reflecting weight change, with the full synaptic operation completing in 305 ns (see Fig.\ref{fig_sim}(b)).The applied current density ia J = 2.5~MA/cm\textsuperscript{2} with pulse width and period of 2.2~ns and 2.5~ns respectively. We observe the minimum conductance ($\mathrm{G_{min}}$) at t = 0 ns and maximum conductance of $\mathrm{G_{max}}$ at t = 142.57 ns. The synaptic weight $W_{i,j}$ is defined as the difference between the skyrmion device conductance and a parallel conductance, $\mathrm{G_{\text{parallel}} = (G_{\text{min}} + G_{\text{max}})/2}$, introduced to support both positive and negative weights. The resulting weights are clipped within $\pm$1 to maintain bounded conductance values. The synapse device requires $\mathrm{E_{write} = 4.23~fJ/state}$ (for $\mathrm{\mathrm{I_{charge}} = 8.3~mA, T_p =~2.5~ns}$) to move the skymions in the nano tracks for neural network implementation. 
\subsubsection{ReLU}
The ReLU outputs the input for positive values and zero otherwise. Figure~\ref{lrelu and leakyrelu design}(h) shows the DW positions for input charge current and their normalized resistances. The Verilog-A is embedded with $\mathrm{I_{charge}}$, time, and resistance, serve as the variable resistor in the divider of Fig.~\ref{lrelu and leakyrelu design}(e). The HSPICE circuit uses a fixed 10.7 K$\Omega$ and a variable resistor (DW readout at T2), feeding a 16 nm predictive technology model~(PTM)) CMOS inverter. The normalized $\mathrm{V_{out}}$ reproduces ReLU behavior over -34.8 $\mu$A to 34.8~$\mu$A  (Fig.~\ref{fig_sim}(d)) for DCGAN architecture. The energy consumption of a single ReLU module is 9.16 fJ.
\subsubsection{Leaky ReLU}
The leaky ReLU also outputs the input (x) for positive value but for negative values it is a non-zero gradient (ax) that mitigates the dying ReLU issue. Figure~\ref{lrelu and leakyrelu design}(j) shows the plot obtained from the parallel MTJ readout , the normalized conductance along with the DW position snapshots for the applied input charge current. The DW at $\mathrm{I_{charge}}$ = 0 $\mu$A is at position 37.6364 nm, K\textsubscript{u} = $\mathrm{0.60~MJ/m^3}$ settling in 10 ns. At positive applied current, the domain wall moves under the right MTJ at 83 nm for $\mathrm{I_{charge}}$ = 22.54 $\mu$A. As their is a parallel MTJ readout, a variable conductance is contributed by $\mathrm{MTJ_{right}}$ and a fixed conductance by $\mathrm{MTJ_{left}}$ at terminal T2L. Similarly, for negative $\mathrm{I_{charge}}$ = -23 $\mu$A, DW moves under $\mathrm{MTJ_{left}}$ at 13 nm, and the corresponding $\mathrm{MTJ_{right}}$ will contribute a constant conductance to the final summed conductance read at terminal T2L. While the maximum conductance ($\mathrm{G_{max}}$) = 0.1331~$\mathrm{m\mho}$ is obtained for  $\mathrm{I_{charge}}$ = 34.5 $\mu$A at  K\textsubscript{u} = $\mathrm{0.8539~MJ/m^3}$ with domain wall at 87.8687 nm settling in 0.5 ns. Similarly, the minimum conductance ($\mathrm{G_{min}}$) = 0.1036~$\mathrm{m\mho}$ is obtained at $\mathrm{I_{charge}}$ = -34.5 $\mu$A at  K\textsubscript{u} = $\mathrm{0.7040~MJ/m^3}$ with domain wall at 12.90 nm settling in 0.2 ns. DW annihilates for -41.4 $\mu$A $>$ $\mathrm{I_{charge}}$ or $\mathrm{I_{charge}}$ $>$ 37.72 $\mu$A. The domain wall gets reset in 0.2 ns to 10 ns and settles in 0.2 ns to 9.83 ns. Similar to ReLU, In leaky ReLU also the obtained resistance values are simulated as variable resistor in the voltage divider circuit as shown in Fig.~\ref{lrelu and leakyrelu design}(f). The information is embedded in verilog A and the circuit implemented using HSPICE , with a fixed resistor ($R_{1L}$ = 8.865 K$\mho$ and a variable resistor is fed to the CMOS inverter pair via voltage divider . The normalized $V_{outL}$ emulates the leaky ReLU functionality in the range -20 $\mu$A to 20 $\mu$A for further processing in the DCGAN architecture as shown in Fig.~\ref{fig_sim}(e). The alpha = 0.2 (negative axis slope) for leaky ReLU is obtained by controlling the left and right MTJ area. The energy consumed by the single leaky ReLU module is 0.192 pJ.
\begin{figure}[!htbp]
    \centering
    \includegraphics[height=2.05in, width=1\linewidth]{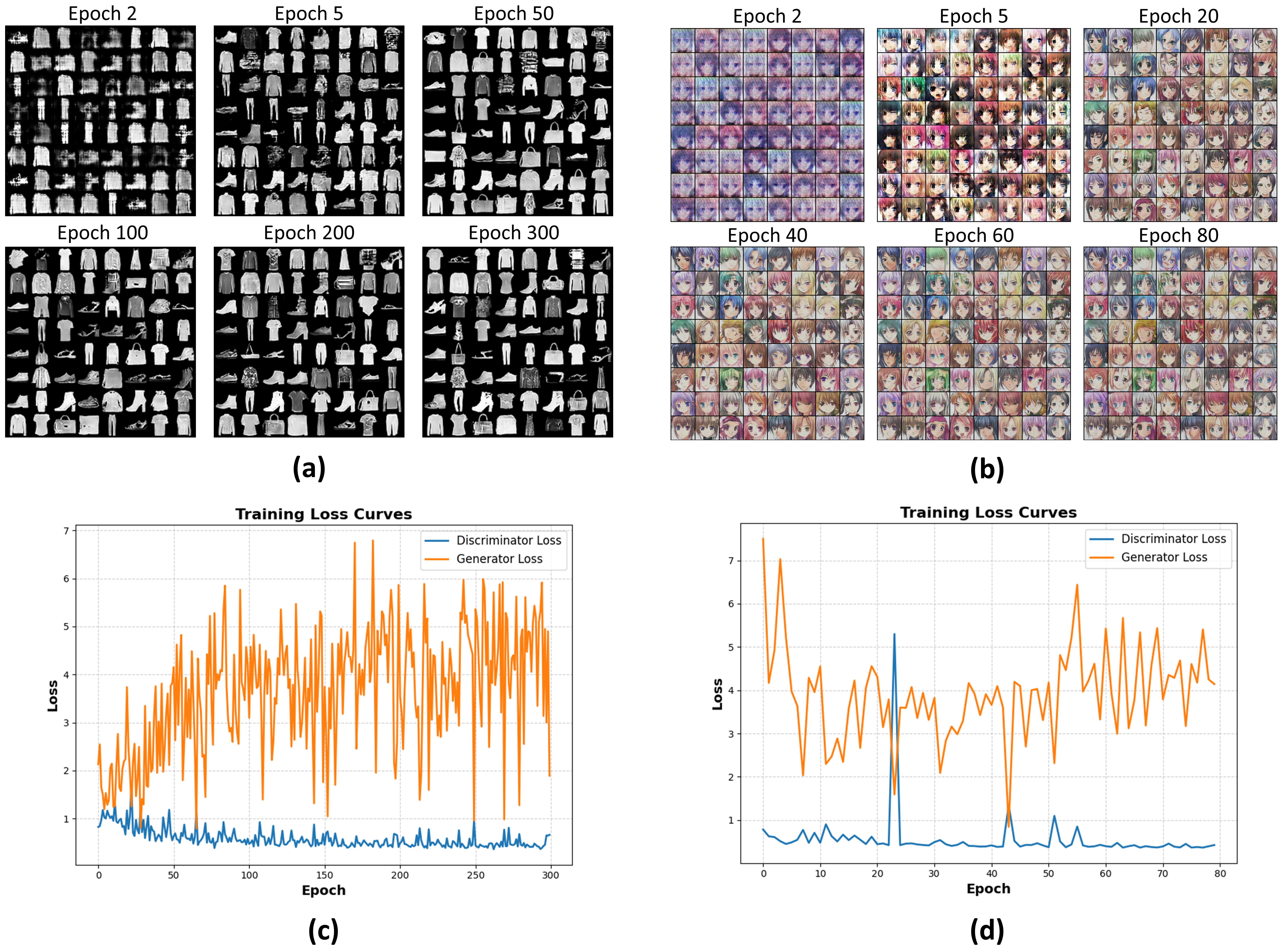}
    \caption{Generated adversarial image samples for (a) Fashion MNIST dataset (b) Anime face dataset. (c) Generator and Discriminator losses per epoch for (c) Fashion MNIST dataset (d) Anime face dataset.}
    \label{lossnepoch}
\end{figure}
\begin{table*}[!ht]
\centering
\caption{Energy analysis for spintronic DCGAN training}
\label{tab:energy_timing}
\resizebox{\linewidth}{!}{%
\begin{tabular}{lcccccccc}
\hline
\textbf{Dataset} & \textbf{$E_{\text{Forward}}$} & \textbf{$E_{\text{Backward}}$} & \textbf{$E_{\text{weight update}}$} & \textbf{$E_{\text{weight update Adam}}$} & \textbf{Training Images} & \textbf{Batches} & \textbf{$E_{\text{Training/Batch}}$} & \textbf{$E_{\text{Training/Image}}$}\\
 & \textbf{(nJ)} & \textbf{(nJ)} & \textbf{(nJ)} & \textbf{(nJ)} & & & \textbf{($\mu$J)} & \textbf{(nJ)} \\ \hline
Fashion MNIST & 4.9 & 9.8 & 11.48 & 34.44 & 60,000 & 493 & 1.916 & 14.97\\
Anime Face & 24.72 & 49.44 & 26.84 & 80.52 & 57,209 & 447 & 9.57 & 74.7 \\ 
\hline
\end{tabular}%
}
\end{table*}
\subsection{Evaluation metrics}\label{Evaluation metrics}
The binary cross entropy  losses for Fashion MNIST and Anime face dataset are shown in Fig.~\ref{lossnepoch}(c,d) estimating qualitatively the stabilization of our proposed DCGAN spintronic architecture. Both generator and discriminator losses saturate after sufficient training iterations. Hyperparameters follow standard DCGAN guidelines and were validated through preliminary stability tests. For Fashion MNIST (28 x 28), stable training was achieved with a batch size of 128 and a symmetric learning rate of 2e-4 for both networks. For the higher resolution Anime Face dataset (64 x 64), visual diversity and stability improved with an asymmetric learning rate setup: 1e-4 (generator) and 2e-4 (discriminator). Final hyperparameters were selected based on smooth loss saturation, consistent convergence across runs, improved image quality and FID during training.\\
\indent For the Anime face dataset in Fig.~\ref{lossnepoch}(d), the generator loss begins relatively high ($>$7), while the discriminator loss is initially low, reflecting generators initial training struggles to produce realistic images and discriminator easily distinguishing between real and fake samples. As training progresses, the generator performance improves, resulting in a gradual decrease in generator loss. After 60 iterations, the generator loss stabilizes between 3 to 6, while the discriminator loss converges tightly around 0.5, indicating a balanced and stable adversarial dynamic.\\
\indent Similarly, for the Fashion MNIST dataset in  Fig.~\ref{lossnepoch}(c), the generator loss begins at 2.5 and the discriminator loss above 1, indicating initial discriminator domination. With training, generator loss gradually increases and saturates between 2 to 6 after 150 iterations. The discriminator also saturates and is eventually fooled by the generator. Figure~\ref{lossnepoch}(a and b) shows the progressive improvements in generated image samples for the datasets. The initial noise evolves into realistic images with learned complexity as training stabilizes with succeeding epochs. The bounded and stable loss against training, confirms the absence of convergence failures\cite{kodali2017convergence}.\\
\indent Mode collapse was evaluated through Fréchet Inception Distance (FID)\cite{yu2021frechet}, which reflects both the visual fidelity and the diversity of generated samples. FID is computed using the PyTorch FID module using Inception-V3 2048 dimensional feature layer. For each dataset, equal numbers of real test images and generated samples were compared. All images were resized to 299 x 299, normalized to the [0,1][0,1][0,1] range, and converted to 8-bit format before feature extraction. Grayscale Fashion-MNIST images were replicated to three channels, whereas Anime Face images were used in RGB format. FID was accumulated over the full test dataloader, following standard FID protocol\cite{yu2021frechet}. The proposed spintronic DCGAN achieves FID of 27.5 for Fashion MNIST\cite{costa2019coegan} and 45.4 for Anime Face dataset\cite{rajarajeswari2025performance}. 
These low FID values indicates high sample diversity and negligible mode collapse, consistent with stable loss curves and the absence of visually repetitive generated samples.\\
\indent We perform hardware aware evaluation by adding Gaussian noise into the latent vector and network weights to emulate device variability or defects and read noise. Synapse weights are clipped to $\pm 1$, matching the skyrmion synapse constraints. Generated images maintain high fidelity and moderate variations for latent noise $\sigma$ of 0.1 to 0.2. Although, weight noise of 2\% to 5\%, significantly increases FID ($\sim$300 for Fashion MNIST and $\sim$180 for Anime face dataset) when applied excessively, but moderate levels improve robustness. Input noise has a smaller impact on FID ($\sim$35 for Fashion MNIST and $\sim$16 for Anime face dataset) than weight noise. Hardware aware weight clipping ensured both reliability and efficiency. Overall, combining moderate latent and weight noise with weight clipping provided a balanced trade-off between image realism, diversity, and hardware compatibility.\\
\indent We assume that the energy consumed by the ReLU and Leaky ReLU activations dominates the forward pass energy, $\mathrm{E_{\text{Forward}}}$, in the DCGAN. The backward pass energy, $\mathrm{E_{\text{Backward}}}$, is taken as twice the forward energy. Each trainable parameter undergoes one update per sample, with an energy cost of $\mathrm{E_{\text{weight update}}}$. Since the Adam optimizer performs three internal state updates per parameter, the effective update energy becomes $\mathrm{E_{\text{weight update adam}}}$. Table~\ref{tab:energy_timing} summarizes these energy parameters along with the total training energy per batch, $\mathrm{E_{\text{Training/Batch}}}$, and the energy required to train a single image, $\mathrm{E_{\text{Train/Image}}}$.\\
\indent Additionally, we evaluated the trained DCGAN under synaptic quantization (4-bit, 5-bit, 6-bit and 8-bit synapses) as loss behavior varies with bit width. Mean squared error (MSE) analysis indicates 4 to 5-bit quantization add only small deviations from floating point weights, while higher bit-widths, preserve weights with near minimal error.\\
\indent From a hardware perspective, lower bit widths substantially reduce storage and read/write energy, improving density and power efficiency. Higher-bits improved fidelity while still lowering memory bandwidth and improving computational throughput compared to full-precision weights. Thus, quantization yields substantial efficiency gains, though very low precision (4-bit) introduces quantization noise that can impact GAN stability.
section{Conclusion}\label{conclusion}
We present a spintronic DCGAN architecture, featuring a compact 6-bit skyrmion based synapse, a hybrid CMOS domain wall ReLU circuit, and our newly proposed Leaky ReLU for energy efficient image generation. Building on our prior skyrmionic synapse and ReLU device  results\cite{gupta2025magnetic}, the new contributions are: 1) A tunable CMOS Leaky ReLU with slope 0.2 obtained using summed conductance readout from a SOT driven DW with piecewise parabolic anisotropy, achieving 0.192 pJ energy; 2) Integration of skyrmionic synapse into the modified generator and discriminator along with DW activations: ReLU and Leaky ReLU units; 3) An end to end spintronic DCGAN across grayscale and colored dataset. Our spintronic DCGAN demonstrates low loss, stable convergence, and high quality image generation, highlighting spintronic devices as promising building blocks for energy efficient generative AI models through device system co-design.
\section*{Author contributions}
A.S. conceptualized the study, S.G. performed the simulations and wrote the manuscript. A. and V.V. contributed to the simulations. S.G.and A.S. revised the manuscript. A.S and B.M. supervised the research, provided funding, and reviewed the manuscript. All authors reviewed and approved the final manuscript.
    \bibliographystyle{IEEEtran}
    \bibliography{Reference}

\end{document}